\documentclass[prb,preprint,showpacs,preprintnumbers,amsmath,amssymb]{revtex4}

\usepackage[dvips]{graphicx}
\usepackage{dcolumn}
\usepackage{bm}
\usepackage{times}
\usepackage{mathptmx}
\usepackage{color}
\begin{document}

\title[]{Incipient ferroelectricity in 2.3\% tensile-strained CaMnO$_3$ films}

\author{T. G\"{u}nter,$^1$ E. Bousquet,$^{2,3}$ A. David,$^4$ Ph.\ Boullay,$^4$ Ph. Ghosez,$^3$ W. Prellier,$^4$ and M. Fiebig$^{1,2}$}

 \affiliation{$^1$HISKP, Universit\"{a}t Bonn, Nussallee 14-16, 53115 Bonn, Germany}
 \affiliation{$^2$Department of Materials, ETH Zurich, Wolfgang-Pauli-Strasse 10, 8093 Zurich, Switzerland}
 \affiliation{$^3$Physique Th\'{e}orique des Mat\'{e}riaux, Universit\'{e} de Li\`{e}ge, All\'{e}e du 6 ao\^{u}t, 17, 4000 Sart Tilman, Belgium}
 \affiliation{$^4$Laboratoire CRISMAT, CNRS UMR 6508, ENSICAEN, 6 Bd.\ Mar\'{e}chal Juin, 14050 Caen Cedex 4, France}

\begin{abstract}
Epitaxial CaMnO$_3$ films grown with 2.3\% tensile strain on (001)-oriented LaAlO$_3$ substrates
are found to be incipiently ferroelectric below 25~K. Optical second harmonic generation (SHG) was
used for the detection of the incipient polarization. The SHG analysis reveals that CaMnO$_3$
crystallites with in-plane orientation of the orthorhombic $b$ axis contribute to an electric
polarization oriented along the orthorhombic $a$ (resp.\ $c$) axis in agreement with the
predictions from density functional calculations.
\end{abstract}


\date{\today}

\maketitle

\section{Introduction: Multiferroics with strain-driven ferroelectricity}

Control of the magnetic response by electric fields and of the dielectric response by magnetic
fields is highly interesting for spintronics applications and devices based on a rigid coupling of
magnetic and dielectric properties. The possibly most fertile source for magnetoelectric cross
correlations are compounds with a coexistence of magnetic and electric long-range order, called
multiferroics.\cite{Eerenstein2006,Cheong2007} The magnetoelectric phase coexistence is inherently
rare: The established mechanisms driving magnetic and ferroelectric order require partially filled
and empty $3d$ orbitals, respectively, and are therefore mutually exclusive unless they arise from
different lattice sites.\cite{Hill2000,Filippetti2002} This lead to an intense search for
alternative mechanisms promoting ferroelectricity in the presence of long-range magnetic order. A
variety of mechanisms were identified up to now.\cite{Khomskii2009} One of the most notable
examples is BiFeO$_3$, the only robust ambient multiferroic, in which ferroelectricity is driven
by an electronic $6s$ lone pair of the Bi$^{3+}$ ions. Furthermore, in compounds where the
magnetic long-range order breaks the inversion symmetry and induces a spontaneous polarization the
magnetoelectric effects are intrinsically strong because of the rigid coupling of the improper
ferroelectric polarization to the proper magnetic order
parameter(s).\cite{Newnham1978,Kimura2003,Hur2004,Cheong2007,Jia2007}

In spite of the impressive magnetoelectric coupling effects observed in some of these compounds an
ambient multiferroic with pronounced and strongly coupled magnetization and polarization has not
been discovered thus far. As a consequence, the ambition to discover fundamentally new mechanisms
promoting multiferroic order is unbroken. In the past, new multiferroics were usually obtained via
the chemical route by synthesizing materials from chemical building blocks with a high potential
to promote multiferroic order. As alternative it was recently suggested to take a structural route
and modify the unit cell parameters of non-multiferroic compounds until they develop the
magnetoelectric phase coexistence. An efficient way to achieve this is strain. The improvement in
thin-film growth technologies in the past years allows the growth of thin films with epitaxial
strains of several percent.\cite{Schlom2007} Such strain can be used for increasing
ferroelectric\cite{Choi2004} and ferromagnetic\cite{Beach1993,Fuchs2008} transition temperatures,
induce ferroelectricity,\cite{Haeni2004} or enhance the magnetization of a
ferromagnet.\cite{Thiele2007} In 2010 it was shown that tensile biaxial strain of 1.1\% pushes
EuTiO$_3$ into a multiferroic state by inducing a spontaneous polarization estimated as
29~$\mu$C/cm$^2$ that coexists with a ferromagnetic phase.\cite{Lee2010a} First-principles density
functional calculations explained this transition by softening of a polar phonon mode driven by
spin-phonon coupling.\cite{Fennie2006}

Aside from this landmark experiment, strain-induced ferroelectricity complementing magnetic order
was also predicted for EuO and CaMnO$_3$. For EuO, epitaxial tensile or compressive strain in the
order of as much as 4\% is required\cite{Bousquet2010} which is difficult to achieve
experimentally. CaMnO$_3$ is a particularly attractive candidate compound. First, the threshold
tensile epitaxial strain required for stabilizing a ferroelectric phase was predicted to be a
moderate 2.1\%.\cite{Bhattacharjee2009} Second, multiferroic CaMnO$_3$ would create an interesting
exception to the aforementioned $3d^0$ rule because in contrast to other multiferroic perovskite
oxides antiferromagnetism and ferroelectricity in CaMnO$_3$ would both be associated to the
Mn$^{4+}$ cation site. At the bulk level, the antiferrodistortive structural instability is
calculated to dominate the ferroelectric instability, so that the ground-state is non-polar.
However, epitaxial tensile strain is expected to promote the polar mode to the extent that the
compound can eventually become spontaneously polarized in thin films. Third, ``chemical strain''
exerted by replacing Ca by Sr$_{0.55}$Ba$_{0.45}$ in bulk crystals already revealed the emergence
of a ferroelectric state.\cite{Sakai2011}

Unfortunately, the experimental verification of ferroelectricity in strained CaMnO$_3$ films
remained an unsolved task thus far. For CaMnO$_3$ a polarization of 4~$\mu$C/cm$^{2}$ was
predicted to emerge (at 0~K) along the orthorhombic $a$ axis for tensile strain of 2.1\% applied in the
orthorhombic $ac$ plane.\cite{Bhattacharjee2009} This value is orders of magnitude larger than the
spontaneous polarization obtained in the magnetically induced
ferroelectrics.\cite{Kimura2003,Hur2004} However, pyroelectric current measurements are spoiled by
leakage currents and by the inefficiency of generating electric poling fields and pyro-currents
within the film plane. A more suitable experimental approach for the verification of in-plane
ferroelectricity is clearly required.

Here we report the presence of an incipient ferroelectric state below 25~K in epitaxial CaMnO$_3$
films subjected to 2.3\% tensile strain. Incipient ferroelectrics are compounds with emerging
electric-dipolar long-range order where the Curie-Weiss temperature $T_{\rm CW}$ is so close to
zero (either $T_{\rm CW}\gtrsim 0$ or $T_{\rm CW}\lesssim 0$) that the long-range-ordered state is
not yet stabilized down to 0~K.\cite{Sakudo1971,Uwe1973,Uwe1976,Chaves1976,Lemanov2002} Incipient ferroelectrics with $T_{\rm CW}\gtrsim 0$ are sometimes
termed ``quantum paraelectrics''\cite{Muller1979,Zhong1996,Lemanov1999} when the suppression of the ferroelectricity is 
related to quantum fluctuations and distinguished from the ``incipient ferroelectrics'' in a
narrower sense. However, here we follow the majority of published work and refrain from making
this distinction. One of the best known incipient ferroelectrics is SrTiO$_3$ ($T_{\rm CW}>0$)\cite{Muller1979,Viana1994} so
that we use it as reference compound to which we compare our CaMnO$_3$ data.

The polarizability of the CaMnO$_3$ films is demonstrated by optical SHG. The direction of the
incipient polarization is derived from symmetry considerations and found to be oriented diagonally
between the principal in-plane axes associated with the cubic perovskite subcell of the
pseudocubically grown films. This orientation agrees with earlier predictions derived from density
functional calculations; here a refined density functional approach backs up our conclusions.

\section{Probing ferroelectricity by SHG}\label{method}

SHG is a well established tool for probing ferroic order in bulk crystals and thin
films.\cite{Fiebig2005,Kordel2009} The nonlinear optical process describes the generation of a
light wave at the frequency 2$\omega$ in a material with $\omega$ as the frequency of the incident
light.\cite{Shen2002} This is described by the expression
$P_i(2\omega)=\varepsilon_0\chi_{ijk}E_j(\omega)E_k(\omega)$. The component $\chi_{ijk}$ of the
nonlinear susceptibility tensor couples $j$ and $k$ polarized contributions of the electric field
$\vec{E}(\omega)$ of the incident light to an $i$ polarized contribution to the polarization
$\vec{P}(2\omega)$ driving the SHG light field. In the electric-dipole approximation $\hat{\chi}$
is a polar tensor so that components $\chi_{ijk} \neq 0$ are obtained in non-centrosymmetric
systems only.\cite{Shen2002} Thus, SHG is well suited for detecting ferroelectric order breaking
the inversion symmetry.\cite{Uesu1995} In contrast to linear optical techniques the ferroelectric
SHG signal emerges free of background at the Curie temperature. SHG is particularly useful for
probing leaky or in-plane ferroelectricity because in contrast to pyro-current measurements the
finite conduction does not interfere with the detection of the spontaneous polarization and
electrodes are not applied. In addition, the degree of freedom of spatial resolution inherent to
optical techniques allows one to probe the spatial distribution of the spontaneous polarization
and, hence, image domains.

\section{Perovskite subcells in pseudocubic samples}\label{samples}

\begin{figure}[htb]
\centering
\includegraphics[keepaspectratio,clip,width=6cm]{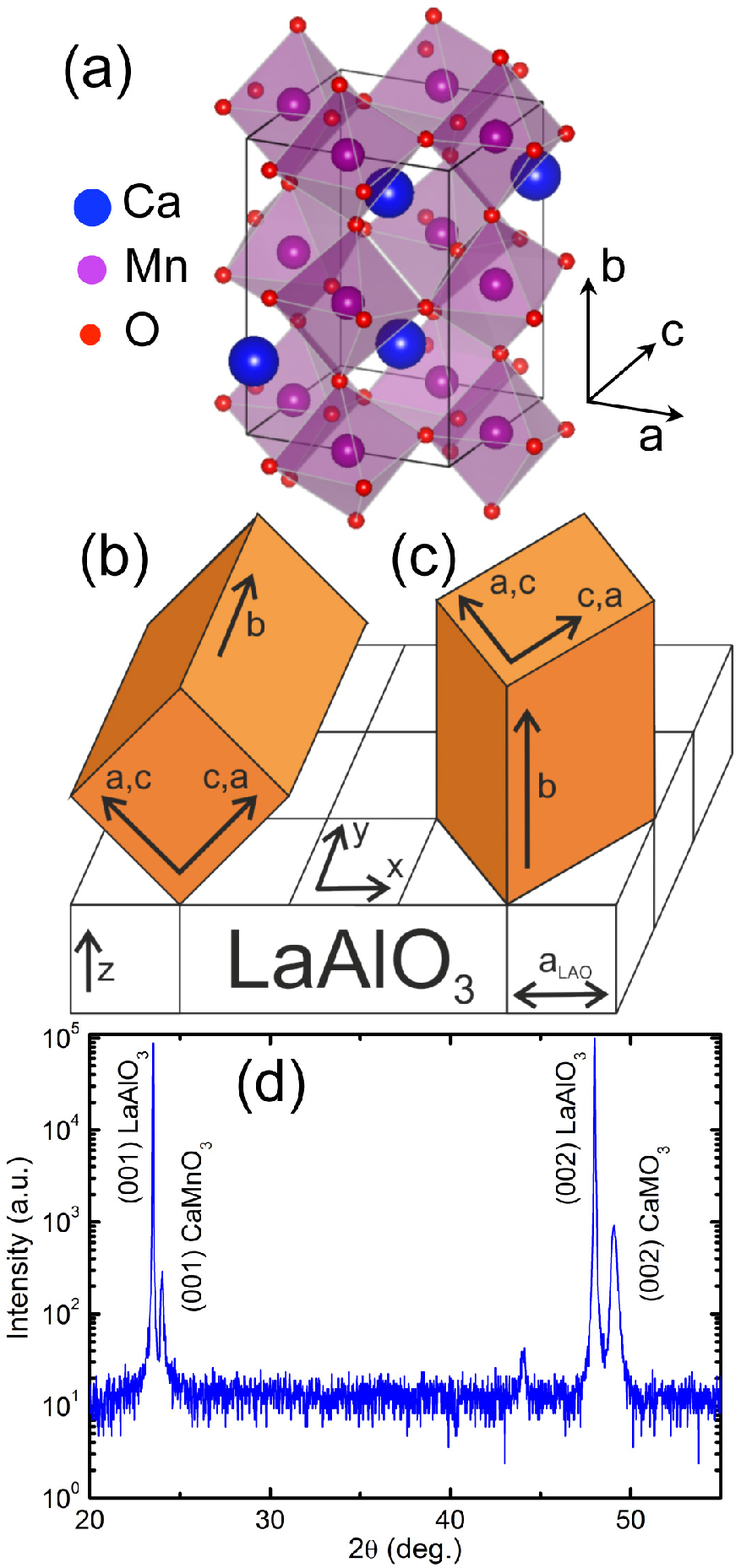}
\caption{Orthorhombic unit cell of CaMnO$_3$ and possible orientations of CaMnO$_3$ grown on
LaAlO$_3$. (a) Orthorhombic unit cell of CaMnO$_3$. (b, c) Principal orientations of CaMnO$_3$
grown on LaAlO$_3$ LaAlO$_3$ with (b) in-plane and (c) out-of-plane orientation of the
orthorhombic $b$ axis. The orthorhombic axes are denoted as $a$, $b$, $c$ whereas the pseudocubic
axes are denoted as $x$, $y$, $z$, with $a_{LAO}$ as the pseudocubic unit cell parameter. (d)
$\theta$-2$\theta$ XRD scan of a typical strained CaMnO$_3$ film epitaxially grown on LaAlO$_3$.
The peak at 44$^{\circ}$ is caused by the sample holder.} \label{fig_orient}
\end{figure}

Bulk CaMnO$_3$ crystallizes in a distorted orthorhombic structure described by the space group
$Pnma$.\cite{Poeppelmeier1982} In orthorhombic coordinates the lattice parameters are $a =
5.279$~\AA, $b = 7.448$~\AA, and $c = 5.264$~\AA. This structure possesses a perovskite subcell
that can be approximated with $a/\sqrt{2}\approx b/2\approx c/\sqrt{2}\approx a_{\rm cube} =
3.72$~\AA{} as cubic lattice parameter as shown in Fig.~\ref{fig_orient}. Below $T_N = 122$~K
CaMnO$_3$ exhibits a G-type antiferromagnetic order. According to
Ref.~\onlinecite{Bhattacharjee2009}, a polar ground state is obtained for tensile strains larger
than 2.1\% for which a cubic substrate with $a_{\rm cube}\geq 3.80$~\AA{} is required.

Our CaMnO$_3$ films (thickness 40~nm) were grown on (001)-oriented LaAlO$_3$ substrates with
$a_{LAO} = 3.81$~\AA{} as lattice constant of the pseudocubic subcell. This corresponds to 2.3\%
tensile strain of the films. Pulsed laser deposition was used for the epitaxial growth of the
CaMnO$_3$ films. The substrates were kept at a constant temperature of 650~$^{\circ}$C during the
deposition, which was carried out at a pressure of 0.04~mbar of flowing oxygen. After the
deposition, the samples were cooled to 400~$^{\circ}$C maintaining the same conditions. The oxygen
pressure was then increased to 300~mbar, followed by slow cooling to room
temperature. The structural study was carried out by x-ray diffraction (XRD) using a Seifert XRD
3000P for the $\theta$-2$\theta$ scans (Cu K$_{\alpha}$, $\lambda = 1.5406$~\AA). The films were
shown to be homogeneous and the structure corresponds to the composition of the target
($\rm{Ca/Mn}=1$) in the limit of accuracy. Sharp and intense diffraction peaks (see
Fig.~\ref{fig_orient}(d)) suggest neatly crystallized single-phase films. Using the XRD results,
the out-of-plane parameter of the films was calculated to be $< 3.71$~\AA, confirming that the
films are under epitaxial tensile strain.

\begin{figure}[htb]
\centering
\includegraphics[keepaspectratio,clip,width=10cm]{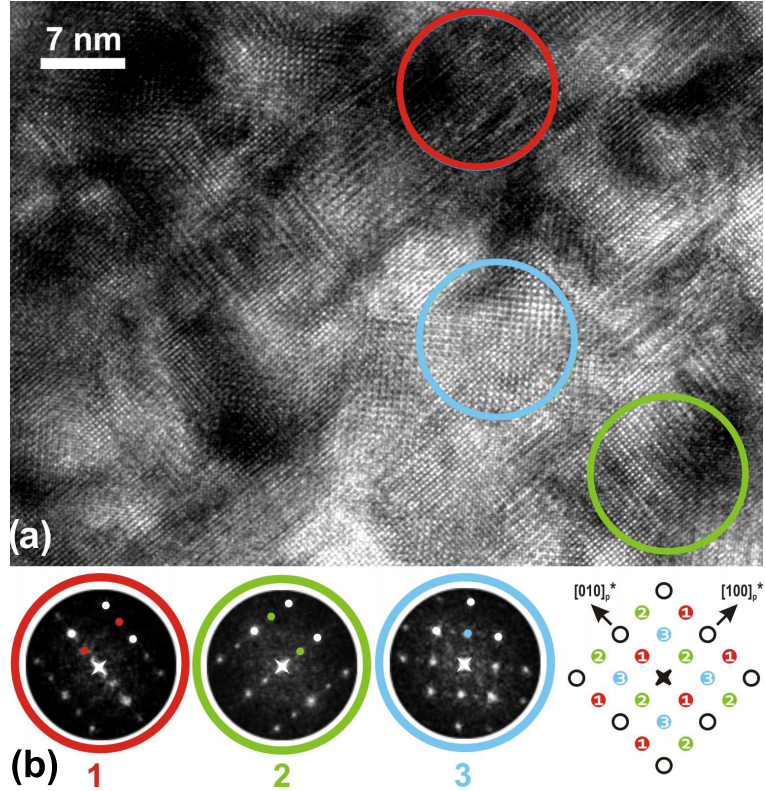}
\caption{(a) High resolution TEM image of a CaMnO$_3$ film (thickness 40~nm) revealing regions
with a different crystallographic orientation and a lateral extension of a few nanometer. On
average, the image possesses a fluctuating contrast suggesting the presence of strain fields due
to the nested configuration of the differently oriented regions.\cite{Rautama2008,Scola2011} (b)
Fourier transformation obtained from three different areas, corresponding to the three types of
orientations of the orthorhombic CaMnO$_3$ unit cell within the film. The schematic drawing
summarizes the four different sets of reflections that can be obtained in the Fourier transform of
the perovskite subcell (white) and the three orientation variants (color). In total six
orientations of the orthorhombic CaMnO$_3$ unit cell are possible, since $a$ and $c$ axis may be
exchanged within the pseudocubic approximation.} \label{fig_tem}
\end{figure}

Figure~\ref{fig_tem} shows a transmission electron microscopy (TEM) image of one of the CaMnO$_3$
films. The TEM image shows that the film is composed of nested regions with a lateral size of
$\lesssim 10$~nm. Fourier transformation (FT) reveals that there are three different types of
regions that are exemplarily highlighted by colored circles. The corresponding states are
associated to the three orientations of the orthorhombic unit cell of CaMnO$_3$ with respect to
the substrate, i.e., with the orthorhombic $b$ axis pointing along the $x$, $y$, or $z$ direction
of the substrate lattice. Applying this assumption, those spots in the FT data that are uniform
across the sample are related to a simple cubic perovskite cell. The remaining spots can be
explained by applying the distorted CaMnO$_3$ perovskite structure with the same rotations of the
MnO$_6$ octahedra that are present in the bulk. As detailed in Fig.~\ref{fig_tem}(b) the
corresponding three sets of spots are then identified as (h k/2 0), (h/2 k 0) and (h/2 k/2 0).
They are associated to [101] (sets 1 and 2) and [010] (set 3) zone axes patterns of the bulk
orthorhombic $Pnma$ structure with cell parameters $a = \sqrt{2} a_c,b = 2a_ c,c = \sqrt{2} a_c$.
Thus, on the one hand, the substrate enforces its cubic lattice parameters onto the CaMnO$_3$
film, but on the other hand, the CaMnO$_3$ retains the bulk orthorhombic atomic distortion for
each of its three orientation states.

We recall, that within the pseudocubic approximation the orthorhombic $a$ axis and $c$ axis may be
interchanged so that we have a total of six possible orientations for the orthorhombic CaMnO$_3$
cell on the LaAlO$_3$ substrate. However, since the resolution of the TEM experiment does not
allow us to distinguish between the $a$ axis and the $c$ axis, only three different orientations
are identified in Fig.~\ref{fig_tem}.

Note that the TEM study indicates the presence of a small amount of secondary phase concluded from
the observation of supplementary spots along the $\langle 110\rangle$ direction of the perovskite
subcell but not detectable in the XRD data. Such superstructure is compatible with the reduced
phases of CaMnO$_3$\cite{Reller1984} and is possibly related to the strained state of the film.

\section{SHG on ferroelectric C\symbol{97}M\symbol{110}O$_3$}

Knowing the appropriate framework for the description of our CaMnO$_3$ films, i.e.\ the
pseudocubic approximation, we can now derive the possible symmetries of the ferroelectric phase
and the resulting polarization selection rules for SHG.

Non-polar CaMnO$_3$ is centrosymmetric with $mmm$ as orthorhombic point symmetry. In the
pseudocubic approximation we neglect the difference between the $a$ and the $c$ axis which changes
the point symmetry to $4/mmm$ with the $b$ axis as the fourfold axis. The spontaneous polarization
of the strained pseudocubic unit cell may be oriented parallel (case i) or perpendicular to the $b$ axis.
In the latter case it may be oriented along the principal $a$ or $c$ axis (case ii) or diagonally,
including an angle of 45$^{\circ}$ with these axes (case iii). Cases (i) and (ii, iii) reduce the
point symmetry to $4mm$ and $mm2$, respectively, with `4' and `2' indicating the direction of the
spontaneous polarization. Lower symmetries do not have to be considered because they would
correspond to other, unphysically arbitrary directions of the spontaneous polarization.

In Table~\ref{table_shg} the SHG contributions for cases (i) to (iii) are given for all the
possible orientations of the orthorhombic unit cell within the pseudocubic lattice. Only tensor
components $\chi_{ijk}$ that can be addressed with light incident perpendicular to the CaMnO$_3$
film ($k\,\|\,z$) are considered. This excludes all the components with $i$,$j$, or $k=z$ since
this would involve longitudinally polarized light. Note that the net SHG yield obtained from the
CaMnO$_3$ film is a mixture of the SHG contributions for all the orientations of the orthorhombic
unit cell that are possible within the pseudocubic lattice.

\section{Experimental setup}

Prior to our SHG experiments, we performed pyroelectric current measurements on the CaMnO$_3$
films. For this purpose, two gold electrodes with a gap of about 1~mm were grown onto the surface
of the films. However, as in previous experiments the leakiness of the epitaxial CaMnO$_3$ films
in combination with the in-plane geometry spoiled the polarization measurement. This leaves SHG as
approach for the detection of a spontaneous polarization.

In the SHG experiments, frequency-tunable laser pulses of about 130~fs are emitted from an optical
parametric amplifier which is operated at 1~kHz by a Ti:sapphire amplifier system. The SHG data
are taken in the spectral range $2\hbar\omega=1.8-3.0$~eV which covers the lowest ${\rm
O}^{2-}\to{\rm Mn}^{4+}$ charge-transfer excitation and the ${\rm Mn}^{4+}(t_{2g})\to{\rm
Mn}^{4+}(e_g)$ intraband transfer.\cite{Satpathy1996,Zampieri2002,Loshkareva2004} In order to
suppress any surface-induced SHG contributions not coupling to ferroelectric order a
near-normal-incidence reflection geometry with a reflection angle incident to the surface normal of approximately $2^{\circ}$ is employed. The polarization of the
incident laser beam is set by a half-wave plate. Optical filters are chosen to suppress background
light in front of the sample and to separate the SHG wave from the fundamental light behind the
sample. The polarization of the SHG light is analyzed with a Glan-Taylor prism. For further
spectral filtering the signal light is transmitted through a monochromator. It is then detected by
a photomultiplier tube and normalized to a reference SHG signal in order to account for spectral
variations of the fundamental light and of the efficiency of the SHG setup. In addition, the
spectra were normalized by dividing the SHG signal obtained from the sample by the spectrally flat
reference SHG response of a silver mirror. Alternatively, for obtaining spatial resolution, the
samples were imaged onto a liquid-nitrogen-cooled camera chip using a standard telephoto lens with
a resolution of about 25~$\mu$m. A liquid-helium-operated cryostat is used to cool the samples to
temperatures between 4.5 and 300~K. Electric fields were applied to the sample via polished steel
plates with a diameter of about 1~cm in between which the samples were mounted.

\section{Experimental results and discussion}

Prior to the experiments on the CaMnO$_3$ films the SHG response of CaMnO$_3$ and LaAlO$_3$ single
crystals was investigated in order to identify any bulk background contributions. As mentioned
before LaAlO$_3$ is a centrosymmetric insulator with a band gap of 5.6 eV\cite{Lim2002} and a
pseudocubic perovskite subcell that exhibits no ferroic order, so that electric-dipole-type SHG
contributions are not expected. In agreement with this, a SHG signal from pure LaAlO$_3$ was not
observed in the temperature range from 5 to 300~K and for photon energies between 1.8 and 3.0~eV.
A CaMnO$_3$ single crystal grown by the floating-zone method revealed a spectrally flat,
temperature-independent SHG signal that was recorded between 4.5 and 150~K. It may be related to
defects, surface contributions, or SHG contributions beyond the electric-dipole approximation.

\begin{figure}[htb]
\centering
\includegraphics[keepaspectratio,clip,width=6cm]{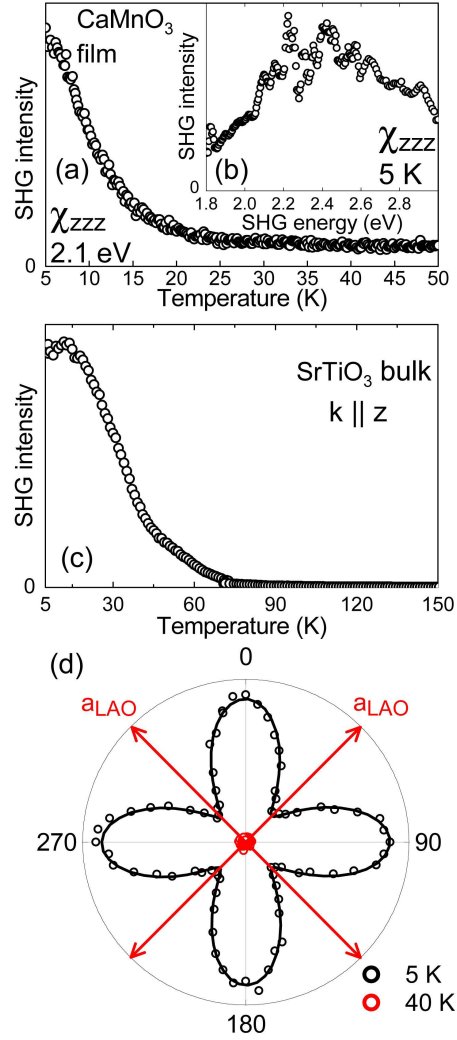}
\caption{(a) Temperature dependence of the SHG signal at 2.1~eV. At 25~K a pronounced polarized
SHG signal emerges and increases continuously towards 5~K. Above 25~K we find a temperature
independent, spectrally featureless SHG background that is present up to at least 150~K. (b) SHG
spectrum of $\chi_{zzz}$ at 5~K. The resonance at 2.4 eV is probably related to
the $t_{2g}\to e_g$ transition of the Mn$^{4+}$ ion which is also observed in x-ray absorption
spectra of bulk CaMnO$_3$.\protect\cite{Zampieri2002} (c) Temperature dependence of the SHG signal
obtained from incipiently ferroelectric bulk SrTiO$_3$ for comparison. The similarity to the
temperature dependence of SHG from CaMnO$_3$ is striking. (d) Polarization dependence of the SHG
signal at 5~K and 40~K. The fit entered as a solid line is perfectly described by a single SHG
tensor component: $\chi_{\rho\rho\rho}=\chi_{\sigma\sigma\sigma}$ (see Table~\ref{table_shg}).}
\label{fig_data}
\end{figure}

Figure \ref{fig_data} shows the normalized spectral, temperature, and polarization dependence of
the SHG signal obtained on the epitaxial CaMnO$_3$ films. The pronounced spectral dependence of
the SHG signal at 5~K demonstrates that it is not related to the background contribution observed
on the CaMnO$_3$ bulk crystals. The spectral dependence is independent of the tensor component. A
peak of the SHG intensity is observed at around 2.4~eV, which is probably related to the
aforementioned $t_{2g}\to e_g$ transition of the Mn$^{4+}$
ion.\cite{Satpathy1996,Zampieri2002,Loshkareva2004} For the following experiments a SHG energy of
2.1~eV was chosen because of the high intensity of the fundamental laser beam in combination with
a reasonably large SHG yield.

Figure \ref{fig_data}(a) shows the temperature dependence of the SHG signal at 2.1~eV. Above 25~K
a constant SHG background is obtained that shows an isotropic polarization dependence and is
present up to at least 150~K. This is the background signal already observed on the bulk sample
and can therefore be regarded as ``zero bias'' of the SHG measurement. At 25~K a pronounced SHG
signal emerges and increases continuously towards 5~K. As Fig.~\ref{fig_data}(d) reveals, the SHG
signal of CaMnO$_3$ has a distinct polarization dependence. The anisotropy measurement shows the
SHG yield polarized parallel to the polarization of the incident fundamental light while rotating
this polarization by $360^{\circ}$. This leads to four equally long lobes with maximum SHG
intensity along the $\langle 110\rangle$ in-plane-diagonal directions of the pseudocubic lattice.
A fit entered as solid line into Fig.~\ref{fig_data}(d) reveals that the polarization dependence
of the SHG signal is perfectly described by a single SHG tensor component:
$\chi_{\rho\rho\rho}=\chi_{\sigma\sigma\sigma}$ (see Table~\ref{table_shg}). This uniquely points
to CaMnO$_3$ crystallites with an out-of-plane orientation of the orthorhombic $b$ axis and a
spontaneous polarization along the orthorhombic $a$ (resp.\ $c$) axis, see
Fig.~\ref{fig_orient}(c). With two possible in-plane orientations of this polarization the SHG
signal from crystallites of either orientation adds up to reveal a fourfold SHG anisotropy in
spite of the $mm2$ symmetry of the individual crystallites. This conclusion is in perfect
agreement with the predictions made by Bhattacharjee {\it et al.}\cite{Bhattacharjee2009} and in
Section~\ref{th-outofplane}. In both cases, density functional theory (DFT) is used to consider an
out-of-plane orientation of the orthorhombic $b$ axis, and a spontaneous in-plane polarization
along the orthorhombic $a$ (resp.\ $c$) axis is found.

Hence, both the SHG data in Fig.~\ref{fig_data} and the DFT results point to an electric
polarization in the pseudocubic epitaxial CaMnO$_3$ films that is induced by tensile lattice
strain of 2.3\%. In contrast to the bulk CaMnO$_3$ ground state, the polar mode becomes unstable
in the strained film so that a polar displacement can appear. Note that the emergence of the SHG
signal below 25~K cannot be due to the magnetic order. First, magnetization-induced SHG also be
observed on the CaMnO$_3$ bulk sample, and its polarization would have to be different from the
polarization of the SHG signal in Fig.~\ref{fig_data}(d). Second, DFT predicted that the
antiferromagnetic order and its critical temperature of $T_N=122$~K are not affected by the
substrate strain and the resulting ferroelectric transition.\cite{Bhattacharjee2009} We thus
conclude that strained pseudocubic CaMnO$_3$ constitutes a strain-driven multiferroic below 25~K.

\begin{figure}[htb]
\centering
\includegraphics[keepaspectratio,clip,width=10cm]{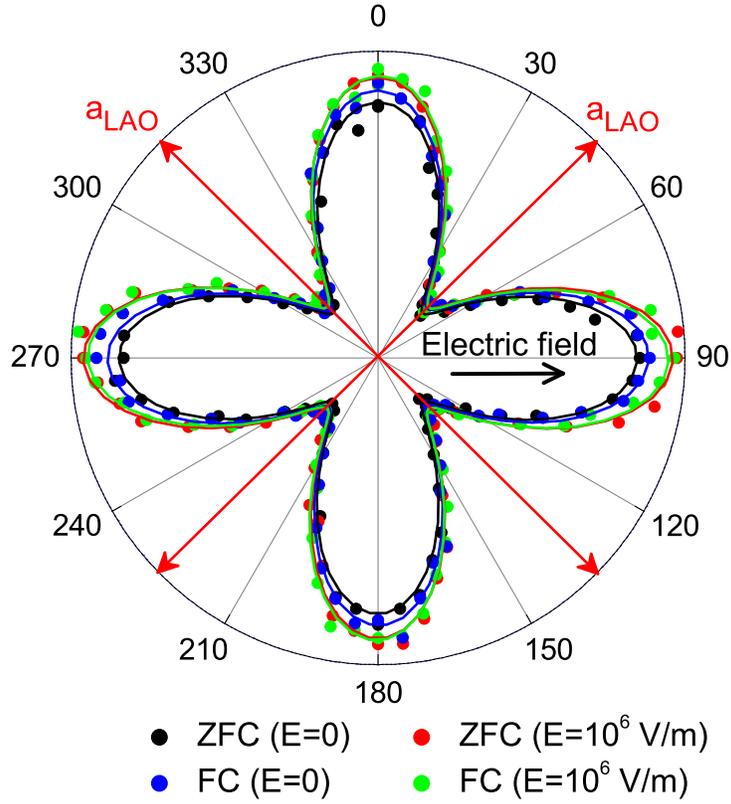}
\caption{Electric-field dependence of the SHG anisotropy. Samples were zero-field cooled (ZFC,
$E=0$) or field cooled (FC, $E=10^6$~V/m) from 40~K and measured at 5~K and 2.1~eV with or without
the electric field applied. The insensitivity of the SHG signal to the electric field indicates
that no polarization switching occurs. Lines are fits to the data according to
Fig.~\protect\ref{fig_data}(c).} \label{fig_field}
\end{figure}

\begin{figure}[htb]
\centering
\includegraphics[keepaspectratio,clip,width=10cm]{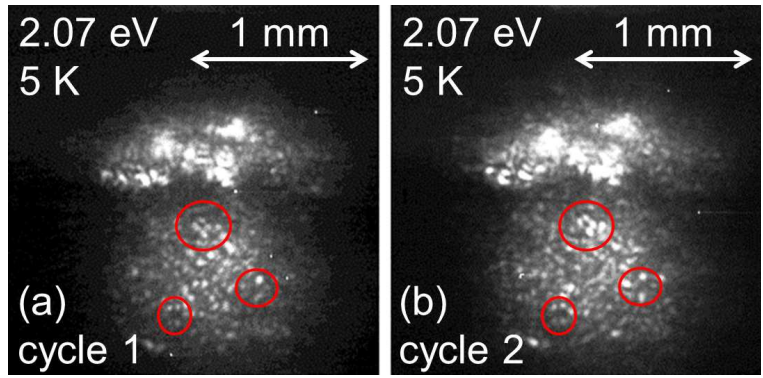}
\caption{Spatially resolved SHG intensity of a CaMnO$_3$ film at 5~K and 2.07~eV SHG photon
energy. A grainy distribution of SHG intensity in the form of resolution-limited speckles is
obtained. The relative position and brightness of the speckles does not change after two
consecutive cooling cycles through 60 K. The red circles mark exemplary areas.}
\label{fig_speckle}
\end{figure}

In order to investigate the rigidity of the polar state, its response to electric poling fields
and thermal cycling was investigated with the results shown in Figs.~\ref{fig_field} and
\ref{fig_speckle}. Figure~\ref{fig_field} shows the effect of a static electric field on the
polarization dependence of the SHG signal. The anisotropy of the SHG signal in the absence of an
electric field and in a static field of $10^6$~V/m during and/or after cooling from 40 to 5~K is
compared. The electric field was applied along as well as diagonally in between the direction
of the pseudocubic axes. (Only the latter case is depicted here since both cases lead to the same
result.) According to Fig.~\ref{fig_field}, the electric field has no effect on the SHG yield. In
addition, cycling the electric field between $\pm 10^6$~V/m showed no sign of polarization
reversal or a hysteresis.

Applying an electric field is expected to reverse the polarization of the domains with a
polarization component antiparallel to the field and, thus, reduce the number of oppositely
polarized domains, ideally towards a single-domain state. Since the SHG signal from oppositely
polarized domains interferes destructively,\cite{Fiebig2002} the field poling is therefore
expected to \textit{enhance} the SHG yield. Yet, the insensitivity of the SHG signal to the
electric field shows that polarization switching does not occur. This may indicate that the
polarized regions are so strongly pinned by the substrate strain and the pseudocubic twinning that
the applied field is too low for polarization reversal. Yet, it is unlikely that even in a
field-cooling experiment no tendency at all for the alignment of the spontaneous polarization
along the applied field is observed, in particular in view of the large value of the spontaneous
polarization (and the related field energy) expected from DFT (Ref.~\onlinecite{Bhattacharjee2009}
and Section~\ref{th-outofplane}).

The spatially resolved SHG intensity of the sample is shown in Fig.~\ref{fig_speckle}. The images
reveal a grainy distribution of the SHG intensity in the form of resolution-limited speckles.
According to Fig.~\ref{fig_speckle}(b) the position and relative brightness of the speckles does
not change when a consecutive annealing cycle through 60~K is applied. (Note that the arrangement
of the speckles in the exemplary red circles is the same in Figs.~\ref{fig_speckle}(a) and
\ref{fig_speckle}(b).) The result does not change when an electric field is applied during the
temperature cycle.

Figure~\ref{fig_speckle} thus reveals that the insensitivity to electric-field cycling is matched
by the insensitivity to thermal cycling. Grainy distribution of the SHG intensity are a result of
the interference of SHG contributions from areas with a size below the optical resolution limit.
Here, these areas can either correspond to the differently oriented crystallographic regions
constituting the pseudocubic structure in Fig.~\ref{fig_tem} or to a distribution of
nanometer-sized ferroelectric domains. The similarity of Figs.~\ref{fig_speckle}(a) and
\ref{fig_speckle}(b) points towards the former. Domains would change in the course of an annealing
cycle unless they are strongly pinned. However, as argued before, pinning effects that are
pronounced enough to withstand an electric poling field when crossing the Curie temperature in the
course of a temperature cycle are unlikely.

We therefore conclude that the polarization picked up by the SHG signal is related to
\textit{incipient} ferroelectricity in the strained CaMnO$_3$ films. As we will see in the
following this conclusion is supported by both experiment and theory. (i) In
Fig.~\ref{fig_data}(c) we show the temperature dependence of the SHG signal from SrTiO$_3$ that we
measured for comparison. The retarded emergence of the signal in this incipient ferroelectric is
qualitatively very similar to the temperature dependence of the SHG signal from CaMnO$_3$. It is
therefore obvious to expect incipient ferroelectricity to be also present in CaMnO$_3$. (ii) The
SHG signal in SrTiO$_3$ displays a distinct spectral and polarization dependence (not shown) just
like the SHG signal in CaMnO$_3$. (iii) As confirmed by Fig.~\ref{fig_data}(c) incipient
ferroelectrics can display a pronounced SHG signal although there is no spontaneous polarization
breaking the inversion symmetry. (We assume that with the high polarizability characteristic for
incipient ferroelectrics the fundamental light wave itself drives the AC polarization that is
breaking the inversion symmetry and probed by SHG. This was also proposed for explaining forbidden
Raman lines in KTaO$_3$.\cite{Akbarzadeh2004}) In the absence of magnetic or structural
transitions in our CaMnO$_3$ films at 25~K, incipient ferroelectricity is thus a compelling reason
for the emergence of the SHG signal. (iv) The proximity to the critical strain predicted by
DFT\cite{Bhattacharjee2009} is in conformity with a ferroelectric potential that is too shallow
to stabilize a spontaneous polarization.

\section{Verification by density functional theory} \label{theory}

As mentioned before the observation of an incipient state with an in-plane
polarization along the orthorhombic $a$ or $c$ axis agrees well with the DFT results reported in
Ref.~\onlinecite{Bhattacharjee2009}. However, in Ref.~\onlinecite{Bhattacharjee2009}
investigations were limited to CaMnO$_3$ films with an out-of-plane orientation of the
orthorhombic $b$ axis as shown in Fig.~\ref{fig_orient}(c). Since regions with an in-plane
orientation of $b$ are present in Fig.~\ref{fig_tem} we have to expand our DFT analysis now. We do
this in two ways: (i) by verifying the results in Ref.~\onlinecite{Bhattacharjee2009} using an
alternative approach for their derivation; (ii) by adding the scenario of an out-of-plane
orientation of $b$. In both cases we retain the pseudocubic approximation of the orthorhombic unit
cell.

\subsection{Out-of-plane orientation of the orthorhombic $b$ axis} \label{th-outofplane}

The formerly used pseudopotential approach with the generalized-gradient-approximation (GGA)
Wu-Cohen exchange correlation functional (as it is implemented in the ABINIT package) was replaced
by the projector-augmented wave (PAW) methodology and the GGA PBEsol\cite{Perdew2008} exchange
correlation functional (as it is implemented in the VASP code.\cite{Kresse1996,Kresse1999})
Converged results were achieved with a plane-wave cutoff of 500~eV and a $k$-point grid of
$4\times 2\times 4$. All calculations were performed with collinear magnetism by taking the G-type
antiferromagnetic order as the magnetic ground state. Phonon frequencies have been calculated for
$Pnma$ space group and different values of epitaxial strain using the frozen-phonon method with
small atomic displacements of $\pm0.03$~\AA. The cell parameter was imposed to be cubic and fixed
to that of the substrate in two directions while it was relaxed in the third direction.

\begin{figure}[htb]
\centering
\includegraphics[keepaspectratio,clip,width=10cm]{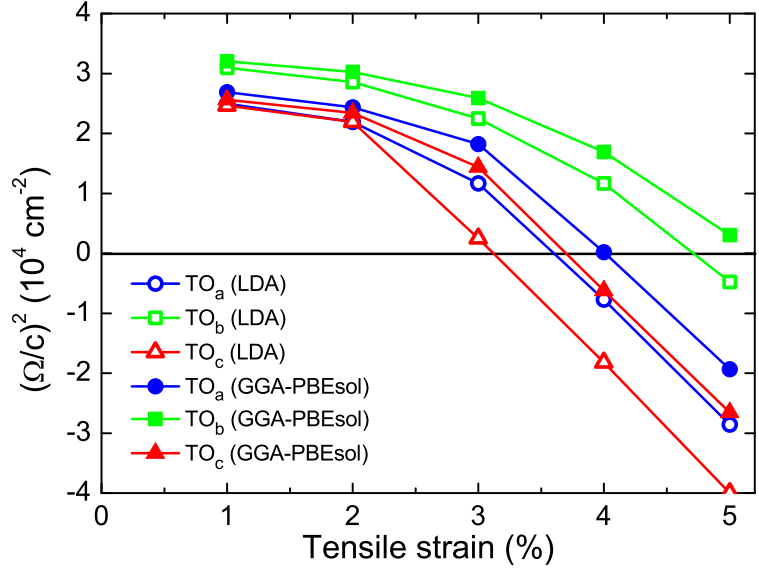}
\caption{Calculated square of the TO frequency $\Omega$ of CaMnO$_3$ versus $a$-$c$ epitaxial
strain. TO$_a$, TO$_b$, and TO$_c$ are the soft TO modes polarized along the orthorhombic $a$,
$b$, and $c$ axis, respectively. At a critical epitaxial strain of 3.2\% (LDA functional) the
TO$_c$ mode becomes unstable and initiates a net polarization along the orthorhombic $c$ axis
(pseudocubic diagonal) reaching 18~$\mu$C/cm$^2$ at an epitaxial strain of 4\%.} \label{fig_th1}
\end{figure}

First, we calculated the evolution of the soft transverse-optical (TO) modes versus the epitaxial
strain applied along the $a$ (resp.\ $c$) direction (see Fig.~\ref{fig_orient}(c)). The strain
dependence of the square of the frequency $\Omega$ of the three soft TO modes (TO$_a$, TO$_b$,
TO$_c$) polarized along the three directions of the crystal is shown in Fig.~\ref{fig_th1}.
Interestingly, all three modes become softer under tensile epitaxial strain. Within the PBEsol
functional the TO$_c$ mode becomes unstable ($\Omega^2<0$) at a critical epitaxial strain of
3.7\%. This is followed by the TO$_a$ mode which becomes unstable at 4.0\%, whereas the TO$_b$
mode remains stable up to 5.0\% , but with a low frequency. This clearly reproduces the fact that
CaMnO$_3$ develops a ferroelectric instability under tensile epitaxial strain that is resulting in
an in-plane polarization. However, we note that the present calculations predict the occurrence of
the ferroelectric instability at a critical tensile epitaxial strain of 3.7\%. This value is
larger than the value of 2.0\% derived previously\cite{Bhattacharjee2009} and refers to the $c$
instead of to the $a$ axis. This difference can be related to the different functionals (GGA
PBEsol instead of GGA Wu-Cohen) and the different DFT methodology (PAW instead of
pseudopotentials) employed here.

In order to scrutinize the influence of the choice of functional, we also performed calculations
by the local density approximation (LDA) based on the PBEsol volume: We froze the volume and the
cell parameters to the values obtained within the PBEsol functional and performed LDA atomic
relaxations and phonon calculations. In Fig.~\ref{fig_th1} the according values of $\Omega^2$ are
shown for the three soft TO modes. We observe the same epitaxial strain sensitivity as reported
with the GGA-PBEsol functional. However, now the ferroelectric instabilities develop at a lower
epitaxial strain: 3.2\% for the TO$_c$ mode, 3.6\% for the TO$_a$ mode, and 4.8\% for the TO$_b$
mode. The result illustrates the considerable influence the choice of the functional has on the
the critical epitaxial strain obtained for the emergence of ferroelectricity. Nevertheless, all
calculations show the same trend, i.e., a ferroelectric $a$- or $c$-axis instability dominates the
response of the CaMnO$_3$ lattice under tensile epitaxial strain applied along the $a$ and $c$
directions.

The ground state was determined by condensing the unstable modes in the structure once they
emerge. We then performed an atomic relaxation for each mode and checked the presence of remaining
instabilities. We thus found that up to a tensile strain of 5.0\% the ground state is always
obtained by the condensation of the TO$_c$ mode alone within the PBEsol and the LDA functionals.
Hence, the condensation of the TO$_c$ mode removes all other instabilities. The ferroelectric
polarization is then predicted to occur along the in-plane $c$ direction. The amplitude of the
polarization depends on the value of the epitaxial strain. With the Berry phase
method\cite{King-Smith1993} we obtain a spontaneous polarization of 12~$\mu$C/cm$^2$ with the
PBEsol functional and 18~$\mu$C/cm$^2$ with the LDA functional for an epitaxial strain of 4.0\%.

Performing the same analysis as in Table II of Ref.~\onlinecite{Bhattacharjee2009} to identify individual atomic contributions to the eigendisplacements of the unstable ferroelectric mode, we also find that the unstable TO$_c$ mode is dominated by Mn and O atomic motions (23\% for Mn, 76\% for O and only 1\% for Ca at an epitaxial strain of 4\% with the PBEsol functional), thus supporting the idea that ferroelectricity in CaMnO$_3$ is mainly driven by the Mn atoms at the B-site, in spite of their partially filled $d$-orbitals.\cite{Bhattacharjee2009}

\subsection{In-plane orientation of the orthorhombic $b$ axis}

\begin{figure}[htb]
\centering
\includegraphics[keepaspectratio,clip,width=10cm]{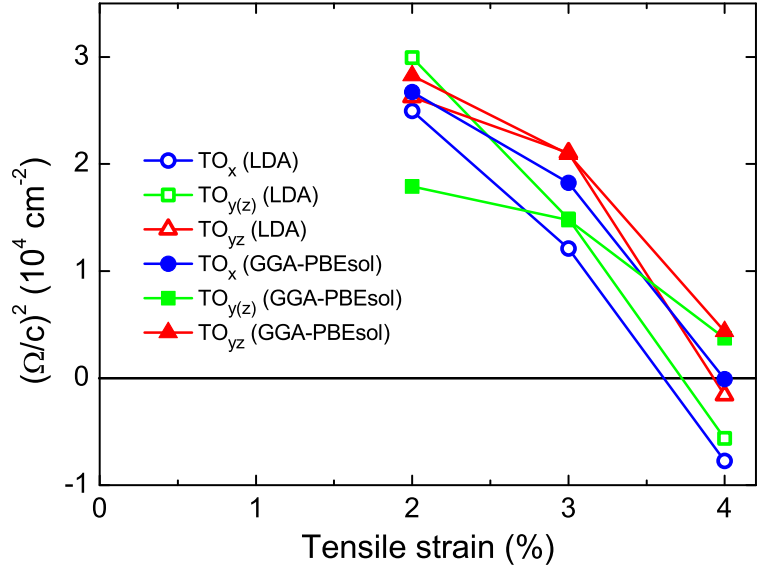}
\caption{Calculated square of the TO frequency $\Omega$ of CaMnO$_3$ versus $b$ epitaxial strain.
The $x$, $y$ and $z$ directions are the directions of the polarization of the modes according to
Fig.~\ref{fig_orient}(b). The TO$_x$ mode becomes unstable at a critical epitaxial strain of 3.6\%
(LDA functional) resulting in a net polarization of 4.6 $\mu$C/cm$^2$ along the $x$ direction of
the crystal (pseudocubic axis) at an epitaxial strain of 4\%.} \label{fig_th2}
\end{figure}

Applying the strain along the orthorhombic $b$ direction (see Fig.~\ref{fig_orient}(b)) constrains
$b$ to the $y$-cell parameter of the substrate. The $a$ and $c$ directions are not parallel to the
substrate which constrains their \textit{projection} to the the $x$-cell parameter of the
substrate. However, the angle between $a$ and $c$ can relax since the out-of-plane projections of
$a$ and $c$ are not constrained.\cite{Eklund2009} For catching this scenario, we performed the
calculations on a cell with 40 atoms. Figure~\ref{fig_th2} shows the evolution of $\Omega^2$ for
the three soft TO modes denoted TO$_x$ (mode polarized along the $x$ direction of the substrate),
TO$_{y(z)}$ (mode polarized along the $y$ direction of the substrate with a small component along
$z$), and TO$_{yz}$ (mode polarized along the $y$ and the $z$ directions of the substrate. The TO
modes become unstable at critical epitaxial strains of, respectively, 3.6\%, 3.8\% and 3.9\%
within the LDA functional. With the PBEsol functional only the TO$_x$ mode becomes unstable at
4.0\% whereas the TO$_{y(z)}$ mode and the TO$_{yz}$ mode are still stable at this strain. In
analogy to the behavior of the TO$_c$ mode with epitaxial strain along $a$ and $c$, condensing the
TO$_x$ mode for strain along $b$ stabilizes the TO$_{yz}$ mode so that the ground state is given
by the condensation of the TO$_x$ mode alone. At an epitaxial strain of 4.0\% this gives rise to a
spontaneous polarization of 4.6~$\mu$C/cm$^2$ along the in-plane $x$ direction with the LDA
functional and a spontaneous polarization of $<0.01$~$\mu$C/cm$^2$ with the PBEsol functional.

While the critical epitaxial strain for inducing a ferroelectric instability depends on the
exchange-correlation functional and the methodology used, all calculations confirm the SHG
results: Independent of the scenario (out-of-plane orientation of $b$ with strain along $a$ and
$c$ or in-plane orientation of $b$ with strain along $b$), a spontaneous polarization along the
orthorhombic $b$ direction of the crystal is never stable. Furthermore, a polarization along the
$x$-direction with in-plane orientation of $b$ is not observed experimentally. This is consistent
with the tendency for the higher values of the epitaxial strain that are required for its
occurrence.

Performing the same analysis as for the case where the strain is applied along the $a$ and $c$ directions, the eigendisplacement of the unstable TO$_x$ mode is also found to be strongly dominated by the Mn and O motions: 45\% for Mn, 54\% for O and only 1\% for Ca at an epitaxial strain of 4\% (PBEsol functional).

\section{Conclusion}

In summary, epitaxial pseudocubically twinned CaMnO$_3$ films with 2.3\% tensile strain were found
to be incipiently ferroelectric below 25~K. The polarizability was detected by optical SHG.
According to the symmetry analysis of the SHG signal CaMnO$_3$ crystallites with out-of-plane
orientation of the orthorhombic $b$ axis contribute to an emerging polarization directed along the
orthorhombic $a$ (resp.\ $c$) axis in agreement with a variety of DFT approaches. The
antiferromagnetic order as well as the incipient ferroelectric order arise from the Mn$^{4+}$
cation site which thus creates a remarkable exception to the $3d^0$ rule for perovskite
multiferroics. With our experiments we emphasize the potential of strained perovskite oxides as
resource for a rich variety of multiferroic compounds. We expect that further research will either
lead to constituents with a lower threshold for strain-induced ferroelectricity or to higher
values of substrate-induced strain stabilizing the ferroelectric state beyond the incipient
behavior found in the present CaMnO$_3$ films.

\section{Acknowledgements}

The work at the Universities of Bonn, Li\`{e}ge, and Caen was supported by the STREP MaCoMuFi
(MP3-CT-2006-033221) of the European Community. The work was additionally supported in Bonn by the SFB 608
of the DFG and in Li\`{e}ge by the EC project OxIDes (NMP3-SL-2008-228989). Authors in Zurich acknowledge support from ETH Zurich. E. B. thanks the FRS-FNRS Belgium for support. M. F. thanks for support by the IMI Program of the National Science
Foundation under Award No.\ DMR-0843934, managed by the International Center for Materials
Research, UC Santa Barbara, USA. Ph. G. thanks the Francqui Foundation for Research Professorship. The authors thank Y.
Tomioka (NIAIST, Tsukuba, Japan) and Y. Tokura (University of Tokyo, Japan) for providing the
CaMnO$_3$ bulk crystal. They further thank N. A. Spaldin for fruitful discussions and L. Gouleuf
for the preparation of the samples used in the TEM experiments.



\clearpage

\begin{table}[htpd]
    \centering
        \begin{tabular}{>{\centering\arraybackslash}p{0.2\columnwidth}>{\centering\arraybackslash}p{0.07\columnwidth}>{\centering\arraybackslash}p{0.07\columnwidth}|>{\centering\arraybackslash}p{0.3\columnwidth}}
        \toprule
\multicolumn{3}{c|}{Pseudocubic direction} & SHG contributions\\
$b$ axis & \multicolumn{2}{c|}{$P_{sp}$} & accessible with $k\,\|\,z$\\ \hline
001 & (i) & 001 & 0\\
& (ii) & 110 & $\rho\rho\rho, \sigma\sigma\rho, \rho\sigma\sigma$ \\
& &1\=10 & $\sigma\sigma\sigma, \rho\rho\sigma, \sigma\rho\rho$ \\
& (iii) & 100 & $xxx, yyx, xyy$\\
& & 010 & $yyy, xxy, yxx$\\ \hline
100 & (i) & 100 & $xxx, yyx, xyy$\\
& (ii) & 011 & $yyy, xxy, yxx$\\
& & 01\=1 & $yyy, xxy, yxx$\\
& (iii) & 010 & $yyy, xxy, yxx$ \\
& & 001 & 0\\ \hline
010 & (i) & 010 & $yyy, xxy, yxx$\\
& (ii) & 101 & $xxx, yyx, xyy$\\
& & 10\=1 & $xxx, yyx, xyy$\\
& (iii) & 100 & $xxx, yyx, xyy$\\
& & 001 & 0\\   \hline \hline
        \end{tabular}
\caption{SHG contributions for all the possible orientations of the orthorhombic b-axis and the
resulting spontaneous polarization $P_{sp}$ within the pseudocubic lattice. Case (i): $P_{sp}$ of
the strained unit cell is oriented parallel to the fourfold $b$-axis. Case (ii): $P_{sp}$ is
perpendicular to the $b$-axis and along the principal $a$ or $c$ axis. Case (iii): $P_{sp}$ is
perpendicular to the $b$-axis, including an angle of 45$^{\circ}$ with $a$ and $c$ (these
directions are denoted as $\sigma$ and $\rho$ with $\sigma\perp\rho$). Only tensor components
$\chi_{ijk}$ that can be addressed with light incident perpendicular to the CaMnO$_3$ film
($k\,\|\,z$) are considered. This excludes all the components with $i$,$j$, or $k=z$ since this
would involve longitudinally polarized light.} \label{table_shg}
\end{table}


\clearpage

\clearpage

\clearpage

\clearpage

\clearpage

\clearpage

\clearpage


\begin{thebibliography}{48}
\expandafter\ifx\csname natexlab\endcsname\relax\def\natexlab#1{#1}\fi
\expandafter\ifx\csname bibnamefont\endcsname\relax
  \def\bibnamefont#1{#1}\fi
\expandafter\ifx\csname bibfnamefont\endcsname\relax
  \def\bibfnamefont#1{#1}\fi
\expandafter\ifx\csname citenamefont\endcsname\relax
  \def\citenamefont#1{#1}\fi
\expandafter\ifx\csname url\endcsname\relax
  \def\url#1{\texttt{#1}}\fi
\expandafter\ifx\csname urlprefix\endcsname\relax\def\urlprefix{URL }\fi
\providecommand{\bibinfo}[2]{#2}
\providecommand{\eprint}[2][]{\url{#2}}

\bibitem[{\citenamefont{Eerenstein et~al.}(2006)\citenamefont{Eerenstein,
  Mathur, and Scott}}]{Eerenstein2006}
\bibinfo{author}{\bibfnamefont{W.}~\bibnamefont{Eerenstein}},
  \bibinfo{author}{\bibfnamefont{N.~D.} \bibnamefont{Mathur}},
  \bibnamefont{and} \bibinfo{author}{\bibfnamefont{J.~F.} \bibnamefont{Scott}},
  \bibinfo{journal}{Nature} \textbf{\bibinfo{volume}{442}},
  \bibinfo{pages}{759} (\bibinfo{year}{2006}).

\bibitem[{\citenamefont{Cheong and Mostovoy}(2007)}]{Cheong2007}
\bibinfo{author}{\bibfnamefont{S.~W.} \bibnamefont{Cheong}} \bibnamefont{and}
  \bibinfo{author}{\bibfnamefont{M.}~\bibnamefont{Mostovoy}},
  \bibinfo{journal}{Nature Materials} \textbf{\bibinfo{volume}{6}},
  \bibinfo{pages}{13} (\bibinfo{year}{2007}).

\bibitem[{\citenamefont{Hill}(2000)}]{Hill2000}
\bibinfo{author}{\bibfnamefont{N.}~\bibnamefont{Hill}}, \bibinfo{journal}{J.
  Phys. Chem. B} \textbf{\bibinfo{volume}{104}}, \bibinfo{pages}{6694}
  (\bibinfo{year}{2000}).

\bibitem[{\citenamefont{Filippetti and Hill}(2002)}]{Filippetti2002}
\bibinfo{author}{\bibfnamefont{A.}~\bibnamefont{Filippetti}} \bibnamefont{and}
  \bibinfo{author}{\bibfnamefont{N.~A.} \bibnamefont{Hill}},
  \bibinfo{journal}{Physical Review B} \textbf{\bibinfo{volume}{65}},
  \bibinfo{pages}{195120} (\bibinfo{year}{2002}).

\bibitem[{\citenamefont{Khomskii}(2009)}]{Khomskii2009}
\bibinfo{author}{\bibfnamefont{D.}~\bibnamefont{Khomskii}},
  \bibinfo{journal}{Physics} \textbf{\bibinfo{volume}{2}}, \bibinfo{pages}{20}
  (\bibinfo{year}{2009}).

\bibitem[{\citenamefont{Newnham et~al.}(1978)\citenamefont{Newnham, Kramer,
  Schulze, and Cross}}]{Newnham1978}
\bibinfo{author}{\bibfnamefont{R.}~\bibnamefont{Newnham}},
  \bibinfo{author}{\bibfnamefont{J.}~\bibnamefont{Kramer}},
  \bibinfo{author}{\bibfnamefont{W.}~\bibnamefont{Schulze}}, \bibnamefont{and}
  \bibinfo{author}{\bibfnamefont{L.}~\bibnamefont{Cross}},
  \bibinfo{journal}{Journal of Applied Physics} \textbf{\bibinfo{volume}{49}},
  \bibinfo{pages}{12} (\bibinfo{year}{1978}).

\bibitem[{\citenamefont{Kimura et~al.}(2003)\citenamefont{Kimura, Goto,
  Shintani, Ishizaka, Arima, and Tokura}}]{Kimura2003}
\bibinfo{author}{\bibfnamefont{T.}~\bibnamefont{Kimura}},
  \bibinfo{author}{\bibfnamefont{T.}~\bibnamefont{Goto}},
  \bibinfo{author}{\bibfnamefont{H.}~\bibnamefont{Shintani}},
  \bibinfo{author}{\bibfnamefont{K.}~\bibnamefont{Ishizaka}},
  \bibinfo{author}{\bibfnamefont{T.}~\bibnamefont{Arima}}, \bibnamefont{and}
  \bibinfo{author}{\bibfnamefont{Y.}~\bibnamefont{Tokura}},
  \bibinfo{journal}{Nature} \textbf{\bibinfo{volume}{406}}, \bibinfo{pages}{55}
  (\bibinfo{year}{2003}).

\bibitem[{\citenamefont{Hur et~al.}(2004)\citenamefont{Hur, Park, Sharma, Ahn,
  Guha, and Cheong}}]{Hur2004}
\bibinfo{author}{\bibfnamefont{N.}~\bibnamefont{Hur}},
  \bibinfo{author}{\bibfnamefont{S.}~\bibnamefont{Park}},
  \bibinfo{author}{\bibfnamefont{P.~A.} \bibnamefont{Sharma}},
  \bibinfo{author}{\bibfnamefont{J.~S.} \bibnamefont{Ahn}},
  \bibinfo{author}{\bibfnamefont{S.}~\bibnamefont{Guha}}, \bibnamefont{and}
  \bibinfo{author}{\bibfnamefont{S.~W.} \bibnamefont{Cheong}},
  \bibinfo{journal}{Nature} \textbf{\bibinfo{volume}{429}},
  \bibinfo{pages}{392} (\bibinfo{year}{2004}).

\bibitem[{\citenamefont{Jia et~al.}(2007)\citenamefont{Jia, Onoda, Nagaosa, and
  Han}}]{Jia2007}
\bibinfo{author}{\bibfnamefont{C.}~\bibnamefont{Jia}},
  \bibinfo{author}{\bibfnamefont{S.}~\bibnamefont{Onoda}},
  \bibinfo{author}{\bibfnamefont{N.}~\bibnamefont{Nagaosa}}, \bibnamefont{and}
  \bibinfo{author}{\bibfnamefont{J.~H.} \bibnamefont{Han}},
  \bibinfo{journal}{Physical Review B} \textbf{\bibinfo{volume}{76}},
  \bibinfo{pages}{144424} (\bibinfo{year}{2007}).

\bibitem[{\citenamefont{Schlom et~al.}(2007)\citenamefont{Schlom, Chen, Eom,
  Rabe, Streiffer, and J.}}]{Schlom2007}
\bibinfo{author}{\bibfnamefont{D.}~\bibnamefont{Schlom}},
  \bibinfo{author}{\bibfnamefont{L.}~\bibnamefont{Chen}},
  \bibinfo{author}{\bibfnamefont{C.}~\bibnamefont{Eom}},
  \bibinfo{author}{\bibfnamefont{K.}~\bibnamefont{Rabe}},
  \bibinfo{author}{\bibfnamefont{S.}~\bibnamefont{Streiffer}},
  \bibnamefont{and} \bibinfo{author}{\bibfnamefont{T.}~\bibnamefont{J.}},
  \bibinfo{journal}{Annual Review of Materials} \textbf{\bibinfo{volume}{37}},
  \bibinfo{pages}{589} (\bibinfo{year}{2007}).

\bibitem[{\citenamefont{Choi et~al.}(2004)\citenamefont{Choi, Biegalski, Y.L.,
  Sharan, Schubert, Uecker, Reiche, Chen, Pan, Gopalan et~al.}}]{Choi2004}
\bibinfo{author}{\bibfnamefont{K.}~\bibnamefont{Choi}},
  \bibinfo{author}{\bibfnamefont{M.}~\bibnamefont{Biegalski}},
  \bibinfo{author}{\bibfnamefont{L.}~\bibnamefont{Y.L.}},
  \bibinfo{author}{\bibfnamefont{A.}~\bibnamefont{Sharan}},
  \bibinfo{author}{\bibfnamefont{J.}~\bibnamefont{Schubert}},
  \bibinfo{author}{\bibfnamefont{R.}~\bibnamefont{Uecker}},
  \bibinfo{author}{\bibfnamefont{P.}~\bibnamefont{Reiche}},
  \bibinfo{author}{\bibfnamefont{Y.}~\bibnamefont{Chen}},
  \bibinfo{author}{\bibfnamefont{X.}~\bibnamefont{Pan}},
  \bibinfo{author}{\bibfnamefont{V.}~\bibnamefont{Gopalan}},
  \bibnamefont{et~al.}, \bibinfo{journal}{Science}
  \textbf{\bibinfo{volume}{306}}, \bibinfo{pages}{1005} (\bibinfo{year}{2004}).

\bibitem[{\citenamefont{Beach et~al.}(1993)\citenamefont{Beach, Borchers,
  Matheny, Erwin, Salamon, Everitt, Pettit, Rhyne, and Flynn}}]{Beach1993}
\bibinfo{author}{\bibfnamefont{R.~S.} \bibnamefont{Beach}},
  \bibinfo{author}{\bibfnamefont{J.~A.} \bibnamefont{Borchers}},
  \bibinfo{author}{\bibfnamefont{A.}~\bibnamefont{Matheny}},
  \bibinfo{author}{\bibfnamefont{R.~W.} \bibnamefont{Erwin}},
  \bibinfo{author}{\bibfnamefont{M.~B.} \bibnamefont{Salamon}},
  \bibinfo{author}{\bibfnamefont{B.}~\bibnamefont{Everitt}},
  \bibinfo{author}{\bibfnamefont{K.}~\bibnamefont{Pettit}},
  \bibinfo{author}{\bibfnamefont{J.~J.} \bibnamefont{Rhyne}}, \bibnamefont{and}
  \bibinfo{author}{\bibfnamefont{C.~P.} \bibnamefont{Flynn}},
  \bibinfo{journal}{Physical Review Letters} \textbf{\bibinfo{volume}{70}},
  \bibinfo{pages}{3502} (\bibinfo{year}{1993}).

\bibitem[{\citenamefont{Fuchs et~al.}(2008)\citenamefont{Fuchs, Arac, Pinta,
  Schuppler, Schneider, and von Loehneysen}}]{Fuchs2008}
\bibinfo{author}{\bibfnamefont{D.}~\bibnamefont{Fuchs}},
  \bibinfo{author}{\bibfnamefont{E.}~\bibnamefont{Arac}},
  \bibinfo{author}{\bibfnamefont{C.}~\bibnamefont{Pinta}},
  \bibinfo{author}{\bibfnamefont{S.}~\bibnamefont{Schuppler}},
  \bibinfo{author}{\bibfnamefont{R.}~\bibnamefont{Schneider}},
  \bibnamefont{and} \bibinfo{author}{\bibfnamefont{H.}~\bibnamefont{von
  Loehneysen}}, \bibinfo{journal}{Physical Review B}
  \textbf{\bibinfo{volume}{77}}, \bibinfo{pages}{014434}
  (\bibinfo{year}{2008}).

\bibitem[{\citenamefont{Haeni et~al.}(2004)\citenamefont{Haeni, Irvin, Chang,
  Uecker, Reiche, Li, Choudhury, Tian, Hawley, Craigo et~al.}}]{Haeni2004}
\bibinfo{author}{\bibfnamefont{J.}~\bibnamefont{Haeni}},
  \bibinfo{author}{\bibfnamefont{P.}~\bibnamefont{Irvin}},
  \bibinfo{author}{\bibfnamefont{W.}~\bibnamefont{Chang}},
  \bibinfo{author}{\bibfnamefont{R.}~\bibnamefont{Uecker}},
  \bibinfo{author}{\bibfnamefont{P.}~\bibnamefont{Reiche}},
  \bibinfo{author}{\bibfnamefont{Y.}~\bibnamefont{Li}},
  \bibinfo{author}{\bibfnamefont{S.}~\bibnamefont{Choudhury}},
  \bibinfo{author}{\bibfnamefont{W.}~\bibnamefont{Tian}},
  \bibinfo{author}{\bibfnamefont{M.}~\bibnamefont{Hawley}},
  \bibinfo{author}{\bibfnamefont{B.}~\bibnamefont{Craigo}},
  \bibnamefont{et~al.}, \bibinfo{journal}{Nature}
  \textbf{\bibinfo{volume}{430}}, \bibinfo{pages}{758} (\bibinfo{year}{2004}).

\bibitem[{\citenamefont{Thiele et~al.}(2007)\citenamefont{Thiele, Doerr,
  Bilani, Rodel, and Schultz}}]{Thiele2007}
\bibinfo{author}{\bibfnamefont{C.}~\bibnamefont{Thiele}},
  \bibinfo{author}{\bibfnamefont{K.}~\bibnamefont{Doerr}},
  \bibinfo{author}{\bibfnamefont{O.}~\bibnamefont{Bilani}},
  \bibinfo{author}{\bibfnamefont{J.}~\bibnamefont{Rodel}}, \bibnamefont{and}
  \bibinfo{author}{\bibfnamefont{L.}~\bibnamefont{Schultz}},
  \bibinfo{journal}{Physical Review B} \textbf{\bibinfo{volume}{75}},
  \bibinfo{pages}{054408} (\bibinfo{year}{2007}).

\bibitem[{\citenamefont{Lee et~al.}(2010)\citenamefont{Lee, Fang, Vlahos, Ke,
  Jung, Kourkoutis, Kim, Ryan, Heeg, Roeckerath et~al.}}]{Lee2010a}
\bibinfo{author}{\bibfnamefont{J.}~\bibnamefont{Lee}},
  \bibinfo{author}{\bibfnamefont{L.}~\bibnamefont{Fang}},
  \bibinfo{author}{\bibfnamefont{E.}~\bibnamefont{Vlahos}},
  \bibinfo{author}{\bibfnamefont{X.}~\bibnamefont{Ke}},
  \bibinfo{author}{\bibfnamefont{Y.}~\bibnamefont{Jung}},
  \bibinfo{author}{\bibfnamefont{L.}~\bibnamefont{Kourkoutis}},
  \bibinfo{author}{\bibfnamefont{J.-W.} \bibnamefont{Kim}},
  \bibinfo{author}{\bibfnamefont{P.}~\bibnamefont{Ryan}},
  \bibinfo{author}{\bibfnamefont{T.}~\bibnamefont{Heeg}},
  \bibinfo{author}{\bibfnamefont{M.}~\bibnamefont{Roeckerath}},
  \bibnamefont{et~al.}, \bibinfo{journal}{Nature}
  \textbf{\bibinfo{volume}{466}}, \bibinfo{pages}{954} (\bibinfo{year}{2010}).

\bibitem[{\citenamefont{Fennie and Rabe}(2006)}]{Fennie2006}
\bibinfo{author}{\bibfnamefont{C.~J.} \bibnamefont{Fennie}} \bibnamefont{and}
  \bibinfo{author}{\bibfnamefont{K.~M.} \bibnamefont{Rabe}},
  \bibinfo{journal}{Physical Review Letters} \textbf{\bibinfo{volume}{97}},
  \bibinfo{pages}{267602} (\bibinfo{year}{2006}).

\bibitem[{\citenamefont{Bousquet et~al.}(2010)\citenamefont{Bousquet, Spaldin,
  and Ghosez}}]{Bousquet2010}
\bibinfo{author}{\bibfnamefont{E.}~\bibnamefont{Bousquet}},
  \bibinfo{author}{\bibfnamefont{N.~A.} \bibnamefont{Spaldin}},
  \bibnamefont{and} \bibinfo{author}{\bibfnamefont{P.}~\bibnamefont{Ghosez}},
  \bibinfo{journal}{Physical Review Letters} \textbf{\bibinfo{volume}{104}},
  \bibinfo{pages}{037601} (\bibinfo{year}{2010}).

\bibitem[{\citenamefont{Bhattacharjee et~al.}(2009)\citenamefont{Bhattacharjee,
  Bousquet, and Ghosez}}]{Bhattacharjee2009}
\bibinfo{author}{\bibfnamefont{S.}~\bibnamefont{Bhattacharjee}},
  \bibinfo{author}{\bibfnamefont{E.}~\bibnamefont{Bousquet}}, \bibnamefont{and}
  \bibinfo{author}{\bibfnamefont{P.}~\bibnamefont{Ghosez}},
  \bibinfo{journal}{Physical Review Letters} \textbf{\bibinfo{volume}{102}},
  \bibinfo{pages}{117602} (\bibinfo{year}{2009}).

\bibitem[{\citenamefont{Sakai et~al.}(2011)\citenamefont{Sakai, Fujioka,
  Fukuda, Okuyama, Hashizume, Kagawa, Nakao, Murakami, Arima, Baron
  et~al.}}]{Sakai2011}
\bibinfo{author}{\bibfnamefont{H.}~\bibnamefont{Sakai}},
  \bibinfo{author}{\bibfnamefont{J.}~\bibnamefont{Fujioka}},
  \bibinfo{author}{\bibfnamefont{T.}~\bibnamefont{Fukuda}},
  \bibinfo{author}{\bibfnamefont{D.}~\bibnamefont{Okuyama}},
  \bibinfo{author}{\bibfnamefont{D.}~\bibnamefont{Hashizume}},
  \bibinfo{author}{\bibfnamefont{F.}~\bibnamefont{Kagawa}},
  \bibinfo{author}{\bibfnamefont{H.}~\bibnamefont{Nakao}},
  \bibinfo{author}{\bibfnamefont{Y.}~\bibnamefont{Murakami}},
  \bibinfo{author}{\bibfnamefont{T.}~\bibnamefont{Arima}},
  \bibinfo{author}{\bibfnamefont{A.}~\bibnamefont{Baron}},
  \bibnamefont{et~al.}, \bibinfo{journal}{Physical Review Letters}
  \textbf{\bibinfo{volume}{107}}, \bibinfo{pages}{137601}
  (\bibinfo{year}{2011}).

\bibitem[{\citenamefont{Sakudo and Unoki}(1971)}]{Sakudo1971}
\bibinfo{author}{\bibfnamefont{T.}~\bibnamefont{Sakudo}} \bibnamefont{and}
  \bibinfo{author}{\bibfnamefont{H.}~\bibnamefont{Unoki}},
  \bibinfo{journal}{Phys. Rev. Lett.} \textbf{\bibinfo{volume}{26}},
  \bibinfo{pages}{851} (\bibinfo{year}{1971}).

\bibitem[{\citenamefont{Uwe et~al.}(1973)\citenamefont{Uwe, Unoki, Fujii, and
  Sakudo}}]{Uwe1973}
\bibinfo{author}{\bibfnamefont{H.}~\bibnamefont{Uwe}},
  \bibinfo{author}{\bibfnamefont{H.}~\bibnamefont{Unoki}},
  \bibinfo{author}{\bibfnamefont{Y.}~\bibnamefont{Fujii}}, \bibnamefont{and}
  \bibinfo{author}{\bibfnamefont{T.}~\bibnamefont{Sakudo}},
  \bibinfo{journal}{Solid State Communications} \textbf{\bibinfo{volume}{13}},
  \bibinfo{pages}{737} (\bibinfo{year}{1973}).

\bibitem[{\citenamefont{Uwe and Sakudo}(1976)}]{Uwe1976}
\bibinfo{author}{\bibfnamefont{H.}~\bibnamefont{Uwe}} \bibnamefont{and}
  \bibinfo{author}{\bibfnamefont{T.}~\bibnamefont{Sakudo}},
  \bibinfo{journal}{Phys. Rev. B} \textbf{\bibinfo{volume}{13}},
  \bibinfo{pages}{271} (\bibinfo{year}{1976}).

\bibitem[{\citenamefont{Chaves et~al.}(1976)\citenamefont{Chaves, Barreto, and
  Ribeiro}}]{Chaves1976}
\bibinfo{author}{\bibfnamefont{A.~S.} \bibnamefont{Chaves}},
  \bibinfo{author}{\bibfnamefont{F.~C.~S.} \bibnamefont{Barreto}},
  \bibnamefont{and} \bibinfo{author}{\bibfnamefont{L.~A.~A.}
  \bibnamefont{Ribeiro}}, \bibinfo{journal}{Phys. Rev. Lett.}
  \textbf{\bibinfo{volume}{37}}, \bibinfo{pages}{618} (\bibinfo{year}{1976}).

\bibitem[{\citenamefont{Lemanov et~al.}(2002)\citenamefont{Lemanov, Sotnikov,
  Smirnova, and Weihnacht}}]{Lemanov2002}
\bibinfo{author}{\bibfnamefont{V.~V.} \bibnamefont{Lemanov}},
  \bibinfo{author}{\bibfnamefont{A.~V.} \bibnamefont{Sotnikov}},
  \bibinfo{author}{\bibfnamefont{E.~P.} \bibnamefont{Smirnova}},
  \bibnamefont{and}
  \bibinfo{author}{\bibfnamefont{M.}~\bibnamefont{Weihnacht}},
  \bibinfo{journal}{Appl.\ Phys.\ Lett.} \textbf{\bibinfo{volume}{81}},
  \bibinfo{pages}{886} (\bibinfo{year}{2002}).

\bibitem[{\citenamefont{M\"{u}ller and Burkard}(1979)}]{Muller1979}
\bibinfo{author}{\bibfnamefont{K.~A.} \bibnamefont{M\"{u}ller}}
  \bibnamefont{and} \bibinfo{author}{\bibfnamefont{H.}~\bibnamefont{Burkard}},
  \bibinfo{journal}{Physical Review B} \textbf{\bibinfo{volume}{19}},
  \bibinfo{pages}{3593} (\bibinfo{year}{1979}).

\bibitem[{\citenamefont{Zhong and Vanderbilt}(1996)}]{Zhong1996}
\bibinfo{author}{\bibfnamefont{W.}~\bibnamefont{Zhong}} \bibnamefont{and}
  \bibinfo{author}{\bibfnamefont{D.}~\bibnamefont{Vanderbilt}},
  \bibinfo{journal}{Phys. Rev. B} \textbf{\bibinfo{volume}{53}},
  \bibinfo{pages}{5047} (\bibinfo{year}{1996}).

\bibitem[{\citenamefont{Lemanov et~al.}(1999)\citenamefont{Lemanov, Sotnikov,
  Smirnova, Weihnacht, and Kunze}}]{Lemanov1999}
\bibinfo{author}{\bibfnamefont{V.~V.} \bibnamefont{Lemanov}},
  \bibinfo{author}{\bibfnamefont{A.~V.} \bibnamefont{Sotnikov}},
  \bibinfo{author}{\bibfnamefont{E.~P.} \bibnamefont{Smirnova}},
  \bibinfo{author}{\bibfnamefont{M.}~\bibnamefont{Weihnacht}},
  \bibnamefont{and} \bibinfo{author}{\bibfnamefont{R.}~\bibnamefont{Kunze}},
  \bibinfo{journal}{Solid State Communications} \textbf{\bibinfo{volume}{110}},
  \bibinfo{pages}{611} (\bibinfo{year}{1999}).

\bibitem[{\citenamefont{Viana et~al.}(1994)\citenamefont{Viana, Lunkenheimer,
  Hemberger, B\"{o}hmer, and Loidl}}]{Viana1994}
\bibinfo{author}{\bibfnamefont{R.}~\bibnamefont{Viana}},
  \bibinfo{author}{\bibfnamefont{P.}~\bibnamefont{Lunkenheimer}},
  \bibinfo{author}{\bibfnamefont{J.}~\bibnamefont{Hemberger}},
  \bibinfo{author}{\bibfnamefont{R.}~\bibnamefont{B\"{o}hmer}},
  \bibnamefont{and} \bibinfo{author}{\bibfnamefont{A.}~\bibnamefont{Loidl}},
  \bibinfo{journal}{Phys. Rev. B} \textbf{\bibinfo{volume}{50}},
  \bibinfo{pages}{601} (\bibinfo{year}{1994}).

\bibitem[{\citenamefont{Fiebig et~al.}(2005)\citenamefont{Fiebig, Pavlov, and
  Pisarev}}]{Fiebig2005}
\bibinfo{author}{\bibfnamefont{M.}~\bibnamefont{Fiebig}},
  \bibinfo{author}{\bibfnamefont{V.}~\bibnamefont{Pavlov}}, \bibnamefont{and}
  \bibinfo{author}{\bibfnamefont{R.}~\bibnamefont{Pisarev}},
  \bibinfo{journal}{Journal of the Optical Society of America B}
  \textbf{\bibinfo{volume}{22}}, \bibinfo{pages}{96} (\bibinfo{year}{2005}).

\bibitem[{\citenamefont{Kordel et~al.}(2009)\citenamefont{Kordel, Wehrenfennig,
  Meier, Lottermoser, Fiebig, Gelard, Dubourdieu, Kim, Schultz, and
  Doerr}}]{Kordel2009}
\bibinfo{author}{\bibfnamefont{T.}~\bibnamefont{Kordel}},
  \bibinfo{author}{\bibfnamefont{C.}~\bibnamefont{Wehrenfennig}},
  \bibinfo{author}{\bibfnamefont{D.}~\bibnamefont{Meier}},
  \bibinfo{author}{\bibfnamefont{T.}~\bibnamefont{Lottermoser}},
  \bibinfo{author}{\bibfnamefont{M.}~\bibnamefont{Fiebig}},
  \bibinfo{author}{\bibfnamefont{I.}~\bibnamefont{Gelard}},
  \bibinfo{author}{\bibfnamefont{C.}~\bibnamefont{Dubourdieu}},
  \bibinfo{author}{\bibfnamefont{J.-W.} \bibnamefont{Kim}},
  \bibinfo{author}{\bibfnamefont{L.}~\bibnamefont{Schultz}}, \bibnamefont{and}
  \bibinfo{author}{\bibfnamefont{K.}~\bibnamefont{Doerr}},
  \bibinfo{journal}{Physical Review B} \textbf{\bibinfo{volume}{80}},
  \bibinfo{pages}{045409} (\bibinfo{year}{2009}).

\bibitem[{\citenamefont{Shen}(2002)}]{Shen2002}
\bibinfo{author}{\bibfnamefont{Y.}~\bibnamefont{Shen}},
  \emph{\bibinfo{title}{The Principles of Nonlinear Optics}}
  (\bibinfo{publisher}{Wiley}, \bibinfo{year}{2002}).

\bibitem[{\citenamefont{Uesu et~al.}(1995)\citenamefont{Uesu, Kurimura, and
  Yamamoto}}]{Uesu1995}
\bibinfo{author}{\bibfnamefont{Y.}~\bibnamefont{Uesu}},
  \bibinfo{author}{\bibfnamefont{S.}~\bibnamefont{Kurimura}}, \bibnamefont{and}
  \bibinfo{author}{\bibfnamefont{Y.}~\bibnamefont{Yamamoto}},
  \bibinfo{journal}{Applied Physics Letters} \textbf{\bibinfo{volume}{66}},
  \bibinfo{pages}{2165} (\bibinfo{year}{1995}).

\bibitem[{\citenamefont{Poeppelmeier et~al.}(1982)\citenamefont{Poeppelmeier,
  Leonowicz, Scanlon, and Longo}}]{Poeppelmeier1982}
\bibinfo{author}{\bibfnamefont{K.}~\bibnamefont{Poeppelmeier}},
  \bibinfo{author}{\bibfnamefont{M.}~\bibnamefont{Leonowicz}},
  \bibinfo{author}{\bibfnamefont{J.}~\bibnamefont{Scanlon}}, \bibnamefont{and}
  \bibinfo{author}{\bibfnamefont{J.}~\bibnamefont{Longo}},
  \bibinfo{journal}{Journal of Solid State Chemistry}
  \textbf{\bibinfo{volume}{45}}, \bibinfo{pages}{71} (\bibinfo{year}{1982}).

\bibitem[{\citenamefont{Reller et~al.}(1984)\citenamefont{Reller, Thomas,
  Jefferson, and Uppal}}]{Reller1984}
\bibinfo{author}{\bibfnamefont{A.}~\bibnamefont{Reller}},
  \bibinfo{author}{\bibfnamefont{J.}~\bibnamefont{Thomas}},
  \bibinfo{author}{\bibfnamefont{D.}~\bibnamefont{Jefferson}},
  \bibnamefont{and} \bibinfo{author}{\bibfnamefont{M.}~\bibnamefont{Uppal}},
  \bibinfo{journal}{Proceedings of the Royal Society London Series A}
  \textbf{\bibinfo{volume}{394}}, \bibinfo{pages}{223} (\bibinfo{year}{1984}).

\bibitem[{\citenamefont{Satpathy et~al.}(1996)\citenamefont{Satpathy,
  Popovi\`{c}, and Vukajlovi\`{c}}}]{Satpathy1996}
\bibinfo{author}{\bibfnamefont{S.}~\bibnamefont{Satpathy}},
  \bibinfo{author}{\bibfnamefont{Z.~S.} \bibnamefont{Popovi\`{c}}},
  \bibnamefont{and} \bibinfo{author}{\bibfnamefont{F.~R.}
  \bibnamefont{Vukajlovi\`{c}}}, \bibinfo{journal}{Physical Review Letters}
  \textbf{\bibinfo{volume}{76}}, \bibinfo{pages}{960} (\bibinfo{year}{1996}).

\bibitem[{\citenamefont{Zampieri et~al.}(2002)\citenamefont{Zampieri, Abbatec,
  Pradoa, Caneiroa, and Morikawad}}]{Zampieri2002}
\bibinfo{author}{\bibfnamefont{G.}~\bibnamefont{Zampieri}},
  \bibinfo{author}{\bibfnamefont{M.}~\bibnamefont{Abbatec}},
  \bibinfo{author}{\bibfnamefont{F.}~\bibnamefont{Pradoa}},
  \bibinfo{author}{\bibfnamefont{A.}~\bibnamefont{Caneiroa}}, \bibnamefont{and}
  \bibinfo{author}{\bibfnamefont{E.}~\bibnamefont{Morikawad}},
  \bibinfo{journal}{Physica B: Condensed Matter}
  \textbf{\bibinfo{volume}{320}}, \bibinfo{pages}{51} (\bibinfo{year}{2002}).

\bibitem[{\citenamefont{Loshkareva et~al.}(2004)\citenamefont{Loshkareva,
  Nomerovannaya, Mostovshchikova, Makhnev, Sukhorukov, Solin, Arbuzova, Naumov,
  Kostromitina, Balbashov et~al.}}]{Loshkareva2004}
\bibinfo{author}{\bibfnamefont{N.~N.} \bibnamefont{Loshkareva}},
  \bibinfo{author}{\bibfnamefont{L.~V.} \bibnamefont{Nomerovannaya}},
  \bibinfo{author}{\bibfnamefont{E.~V.} \bibnamefont{Mostovshchikova}},
  \bibinfo{author}{\bibfnamefont{A.~A.} \bibnamefont{Makhnev}},
  \bibinfo{author}{\bibfnamefont{Y.~P.} \bibnamefont{Sukhorukov}},
  \bibinfo{author}{\bibfnamefont{N.~I.} \bibnamefont{Solin}},
  \bibinfo{author}{\bibfnamefont{T.~I.} \bibnamefont{Arbuzova}},
  \bibinfo{author}{\bibfnamefont{S.~V.} \bibnamefont{Naumov}},
  \bibinfo{author}{\bibfnamefont{N.~V.} \bibnamefont{Kostromitina}},
  \bibinfo{author}{\bibfnamefont{A.~M.} \bibnamefont{Balbashov}},
  \bibnamefont{et~al.}, \bibinfo{journal}{Physical Review B}
  \textbf{\bibinfo{volume}{70}}, \bibinfo{pages}{224406}
  (\bibinfo{year}{2004}).

\bibitem[{\citenamefont{Lim et~al.}(2002)\citenamefont{Lim, Kriventsov,
  Jackson, Haeni, Schlom, Balbashov, Uecker, Reiche, Freeout, and
  Lucovsky}}]{Lim2002}
\bibinfo{author}{\bibfnamefont{S.-G.} \bibnamefont{Lim}},
  \bibinfo{author}{\bibfnamefont{S.}~\bibnamefont{Kriventsov}},
  \bibinfo{author}{\bibfnamefont{T.}~\bibnamefont{Jackson}},
  \bibinfo{author}{\bibfnamefont{J.}~\bibnamefont{Haeni}},
  \bibinfo{author}{\bibfnamefont{D.}~\bibnamefont{Schlom}},
  \bibinfo{author}{\bibfnamefont{A.}~\bibnamefont{Balbashov}},
  \bibinfo{author}{\bibfnamefont{R.}~\bibnamefont{Uecker}},
  \bibinfo{author}{\bibfnamefont{P.}~\bibnamefont{Reiche}},
  \bibinfo{author}{\bibfnamefont{J.}~\bibnamefont{Freeout}}, \bibnamefont{and}
  \bibinfo{author}{\bibfnamefont{G.}~\bibnamefont{Lucovsky}},
  \bibinfo{journal}{Journal of Applied Physics} \textbf{\bibinfo{volume}{91}},
  \bibinfo{pages}{4500} (\bibinfo{year}{2002}).

\bibitem[{\citenamefont{Fiebig et~al.}(2002)\citenamefont{Fiebig, Fr\"{o}hlich,
  Lottermoser, and Maat}}]{Fiebig2002}
\bibinfo{author}{\bibfnamefont{M.}~\bibnamefont{Fiebig}},
  \bibinfo{author}{\bibfnamefont{D.}~\bibnamefont{Fr\"{o}hlich}},
  \bibinfo{author}{\bibfnamefont{T.}~\bibnamefont{Lottermoser}},
  \bibnamefont{and} \bibinfo{author}{\bibfnamefont{M.}~\bibnamefont{Maat}},
  \bibinfo{journal}{Physical Review B} \textbf{\bibinfo{volume}{66}},
  \bibinfo{pages}{144102} (\bibinfo{year}{2002}).

\bibitem[{\citenamefont{Akbarzadeh et~al.}(2004)\citenamefont{Akbarzadeh,
  Bellaiche, Leung, \'{I}\~{n}iguez, and Vanderbilt}}]{Akbarzadeh2004}
\bibinfo{author}{\bibfnamefont{A.~R.} \bibnamefont{Akbarzadeh}},
  \bibinfo{author}{\bibfnamefont{L.}~\bibnamefont{Bellaiche}},
  \bibinfo{author}{\bibfnamefont{K.}~\bibnamefont{Leung}},
  \bibinfo{author}{\bibfnamefont{J.}~\bibnamefont{\'{I}\~{n}iguez}},
  \bibnamefont{and}
  \bibinfo{author}{\bibfnamefont{D.}~\bibnamefont{Vanderbilt}},
  \bibinfo{journal}{Phys. Rev. B} \textbf{\bibinfo{volume}{70}},
  \bibinfo{pages}{054103} (\bibinfo{year}{2004}).

\bibitem[{\citenamefont{Perdew et~al.}(2008)\citenamefont{Perdew, Ruzsinszky,
  Csonka, Vydrov, Scuseria, Constantin, Zhou, and Burke}}]{Perdew2008}
\bibinfo{author}{\bibfnamefont{J.~P.} \bibnamefont{Perdew}},
  \bibinfo{author}{\bibfnamefont{A.}~\bibnamefont{Ruzsinszky}},
  \bibinfo{author}{\bibfnamefont{G.~I.} \bibnamefont{Csonka}},
  \bibinfo{author}{\bibfnamefont{O.~A.} \bibnamefont{Vydrov}},
  \bibinfo{author}{\bibfnamefont{G.~E.} \bibnamefont{Scuseria}},
  \bibinfo{author}{\bibfnamefont{L.~A.} \bibnamefont{Constantin}},
  \bibinfo{author}{\bibfnamefont{X.}~\bibnamefont{Zhou}}, \bibnamefont{and}
  \bibinfo{author}{\bibfnamefont{K.}~\bibnamefont{Burke}},
  \bibinfo{journal}{Physical Review Letters} \textbf{\bibinfo{volume}{100}},
  \bibinfo{pages}{136406} (\bibinfo{year}{2008}).

\bibitem[{\citenamefont{Kresse and Furthm\"{u}ller}(1996)}]{Kresse1996}
\bibinfo{author}{\bibfnamefont{G.}~\bibnamefont{Kresse}} \bibnamefont{and}
  \bibinfo{author}{\bibfnamefont{J.}~\bibnamefont{Furthm\"{u}ller}},
  \bibinfo{journal}{Physical Review B} \textbf{\bibinfo{volume}{54}},
  \bibinfo{pages}{11169} (\bibinfo{year}{1996}).

\bibitem[{\citenamefont{Kresse and Joubert}(1999)}]{Kresse1999}
\bibinfo{author}{\bibfnamefont{G.}~\bibnamefont{Kresse}} \bibnamefont{and}
  \bibinfo{author}{\bibfnamefont{D.}~\bibnamefont{Joubert}},
  \bibinfo{journal}{Physical Review B} \textbf{\bibinfo{volume}{59}},
  \bibinfo{pages}{1758} (\bibinfo{year}{1999}).

\bibitem[{\citenamefont{King-Smith and Vanderbilt}(1993)}]{King-Smith1993}
\bibinfo{author}{\bibfnamefont{R.~D.} \bibnamefont{King-Smith}}
  \bibnamefont{and}
  \bibinfo{author}{\bibfnamefont{D.}~\bibnamefont{Vanderbilt}},
  \bibinfo{journal}{Physical Review B} \textbf{\bibinfo{volume}{47}},
  \bibinfo{pages}{1651} (\bibinfo{year}{1993}).

\bibitem[{\citenamefont{Eklund et~al.}(2009)\citenamefont{Eklund, Fennie, and
  Rabe}}]{Eklund2009}
\bibinfo{author}{\bibfnamefont{C.-J.} \bibnamefont{Eklund}},
  \bibinfo{author}{\bibfnamefont{C.~J.} \bibnamefont{Fennie}},
  \bibnamefont{and} \bibinfo{author}{\bibfnamefont{K.~M.} \bibnamefont{Rabe}},
  \bibinfo{journal}{Physical Review B} \textbf{\bibinfo{volume}{79}},
  \bibinfo{pages}{220101} (\bibinfo{year}{2009}).

\bibitem[{\citenamefont{Rautama et~al.}(2008)\citenamefont{Rautama, Boullay,
  Kundu, Caignaert, Pralong, Karppinen, and Raveau}}]{Rautama2008}
\bibinfo{author}{\bibfnamefont{E.-L.} \bibnamefont{Rautama}},
  \bibinfo{author}{\bibfnamefont{P.}~\bibnamefont{Boullay}},
  \bibinfo{author}{\bibfnamefont{A.}~\bibnamefont{Kundu}},
  \bibinfo{author}{\bibfnamefont{V.}~\bibnamefont{Caignaert}},
  \bibinfo{author}{\bibfnamefont{V.}~\bibnamefont{Pralong}},
  \bibinfo{author}{\bibfnamefont{M.}~\bibnamefont{Karppinen}},
  \bibnamefont{and} \bibinfo{author}{\bibfnamefont{B.}~\bibnamefont{Raveau}},
  \bibinfo{journal}{Chemistry of materials} \textbf{\bibinfo{volume}{20}},
  \bibinfo{pages}{2742} (\bibinfo{year}{2008}).

\bibitem[{\citenamefont{Scola et~al.}(2011)\citenamefont{Scola, Boullay, Noun,
  Popova, Dumont, Fouchet, and Keller}}]{Scola2011}
\bibinfo{author}{\bibfnamefont{J.}~\bibnamefont{Scola}},
  \bibinfo{author}{\bibfnamefont{P.}~\bibnamefont{Boullay}},
  \bibinfo{author}{\bibfnamefont{W.}~\bibnamefont{Noun}},
  \bibinfo{author}{\bibfnamefont{E.}~\bibnamefont{Popova}},
  \bibinfo{author}{\bibfnamefont{Y.}~\bibnamefont{Dumont}},
  \bibinfo{author}{\bibfnamefont{A.}~\bibnamefont{Fouchet}}, \bibnamefont{and}
  \bibinfo{author}{\bibfnamefont{N.}~\bibnamefont{Keller}},
  \bibinfo{journal}{Journal of Applied Physics} \textbf{\bibinfo{volume}{110}},
  \bibinfo{pages}{043928} (\bibinfo{year}{2011}).

\end{thebibliography}
\end{document}